\documentstyle[pra,aps]{revtex}

\newcommand{\beq}{\begin{equation}}
\newcommand{\eeq}{\end{equation}}
\newcommand{\bea}{\begin{eqnarray}}
\newcommand{\eea}{\end{eqnarray}}

\draft

\begin{document}
\title{Cavity implementation of quantum interference in a $\Lambda$-type atom}
\author{Peng Zhou\thanks{
Electronic address: peng.zhou@physics.gatech.edu}}
\address{School of Physics, Georgia Institute of Technology, Atlanta, GA 30332-0430, USA.}
\date{}
\maketitle

\begin{abstract}
A scheme for engineering quantum interference in a $\Lambda$-type atom
coupled to a frequency-tunable, single-mode cavity field with a pre-selected
polarization at finite temperature is proposed. Interference-assisted
population trapping, population inversions and probe gain at one sideband of
the Autler-Townes spectrum are predicted for certain cavity resonant
frequencies.
\end{abstract}


Within recent years, there has been a resurgence of interest in the
phenomenon of quantum interference between different transition paths of
atoms \cite{arim96}. The principal reason is that it lies at the heart of
many new effects and applications of quantum optics, such as lasing without
population inversion \cite{harris}, electromagnetically-induced transparency
\cite{EIT}, enhancement of the index of refraction without absorption \cite
{index}, fluorescence quenching \cite{zhu1,cardimona,zhou1}, spectral line
narrowing \cite{zhou1,nar}.

The basic system consists of a singlet state connected to a closely-spaced
doublet by a single electromagnetic vacuum interaction \cite
{cardimona,zhou1,jan}, so that the two transition pathways from the doublet
states to the singlet are not independent and may interfere. It is important
for these effects that the dipole moments of the transitions involved are
parallel, so that the {\em cross-transition terms} are maximal. From the
experimental point view, however, it is difficult to find isolated atomic
systems which have parallel moments \cite
{harris,cardimona,jan,agarwal,berman}.

Various alternative proposals \cite{EIT,nar,agarwal,scully} have been made
for generating quantum interference effects. For example, for three-level
atomic systems (in $V$, $\Lambda $ and $\Xi $ configurations) excited by two
laser fields: one being a strong pump field to drive two levels (say $
|1\rangle $ and $|2\rangle $) and the other being a weak probe field at
different frequency to probe the levels $|0\rangle $ and $|1\rangle $ or $
|2\rangle $, the strong coherent field can drive the levels $|1\rangle $ and
$|2\rangle $ into superpositions of these states, so that different atomic
transitions are correlated. For such systems, the cross-transition terms are
evident in the atomic dressed picture \cite{EIT,nar,scully}. A four-level
atom with two closely-spaced intermediate states coupled to a two-mode
cavity can also show the effect of quantum interference \cite{agarwal}. In
fact, the experimental observation of the interference-induced suppression
of spontaneous emission was carried out in sodium dimers where the excited
sublevels are superpositions of singlet and triplet states that are mixed by
a spin-orbit interaction \cite{zhu1,berman}.

The major purpose of this Letter is to propose a scheme whereby quantum
interference can be readily engendered in realistic, practical situations.
We study a $\Lambda $-type atom coupled to a frequency-tunable, single-mode
cavity field with a pre-selected polarization which is damped by {\it a
thermal reservoir, }and show that maximal quantum interference
(equivalently, two parallel dipole transition moments) can be achieved in
such a system. Interference-assisted population trapping, population
inversions and probe gain at one component of the Autler-Townes spectrum are
predicted for certain cavity resonant frequencies.

The model consists of a $\Lambda $-type three-level atom with the ground
sublevels $|0\rangle $ and $|1\rangle $, with a level splitting $
\omega_{10}=E_{1}-E_{0}$, coupled by the single-mode cavity field to the
excited level $|2\rangle $. Direct transitions between the ground doublet $
|0\rangle $ and $|1\rangle $ are dipole forbidden. The master equation for
the total density matrix operator $\rho _{T}$ in the frame rotating with the
average atomic transition frequency $\omega _{0}=(\omega_{20}+\omega_{21})/2$
takes the form

\begin{equation}
\dot{\rho}_{T}=-i\left[ H_{A}+H_{C}+H_{I},\,\rho _{T}\right] +{\cal L}\rho
_{T},  \label{master}
\end{equation}
with
\begin{eqnarray}
H_{A} &=&\frac{\omega _{10}}{2}\left( A_{11}-A_{00}\right) , \\
H_{C} &=&\delta \,a^{\dagger }a, \\
H_{I} &=&i\left( g_{1}A_{12}+g_{0}A_{02}\right) a^{\dagger }-h.c., \\
{\cal L}\rho _{T} &=&\kappa (N+1)\left( 2a\rho _{T}a^{\dagger }-a^{\dagger
}a\rho _{T}-\rho _{T}a^{\dagger }a\right)   \nonumber \\
&&+\kappa N\left( 2a^{\dagger }\rho _{T}a-aa^{\dagger }\rho _{T}-\rho
_{T}aa^{\dagger }\right) ,
\end{eqnarray}
where $H_{C}$, $H_{A}$ and $H_{I}$ are the unperturbed cavity, the
unperturbed atom and the cavity-atom interaction Hamiltonians respectively,
while ${\cal L}\rho _{T}$ describes damping of the cavity field by the
continuum electromagnetic modes at finite temperature, characterized by the
decay constant $\kappa $ and the mean number of thermal photons $N$; $a$ and
$a^{\dag }$ are the photon annihilation and creation operators of the cavity
mode, and $A_{ij}=|i\rangle \langle j|$ is the atomic population (the dipole
transition) operator for $i=j$ $(i\neq j)$; $\delta =\omega _{C}-\omega _{0}$
is the cavity detuning from the average atomic transition frequency, and $
g_{i}={\bf e}_{\lambda }\cdot {\bf d}_{i2}\sqrt{\hbar \omega _{C}/2\epsilon
_{0}V}\;(i=0,\,1)$ is the atom-cavity coupling constant with ${\bf d}_{i2}$,
the dipole moment of the atomic transition from $|2\rangle $ to $|i\rangle $
, ${\bf e}_{\lambda }$, the polarization of the cavity mode, and $V$, the
volume of the system. In the remainder of this work we assume that the
polarization of the cavity field is {\em pre-selected}, i.e., the
polarization index $\lambda $ is fixed to one of two possible directions.

In this paper we are interested in the bad cavity limit: $\kappa \gg g_{i}$,
that is the atom-cavity coupling is weak, and the cavity has a low $Q$ so
that the cavity field decay dominates. The cavity field response to the
continuum modes is much faster than that produced by its interaction with
the atom, so that the atom always experiences the cavity mode in the state
induced by the thermal reservoir. Thus one can adiabatically eliminate the
cavity-mode variables, giving rise to a master equation for the atomic
variables only \cite{detail}, which takes the form,
\begin{eqnarray}
\dot{\rho} &=&-i\left[ H_{A},\;\rho \right]  \nonumber \\
&&+\left\{F(-\omega _{10})(N+1)\left[ |g_{0}|^{2}\left( A_{02}\rho
A_{20}-A_{22}\rho \right) +g_{0}g_{1}^{*}A_{02}\rho A_{21}\right] \right.
\nonumber \\
&&+F(\omega _{10})(N+1)\left[ |g_{1}|^{2}\left( A_{12}\rho A_{21}-A_{22}\rho
\right) +g_{0}^{*}g_{1}A_{12}\rho A_{20}\right]  \nonumber \\
&&+F(-\omega _{10})N\left[ |g_{0}|^{2}\left( A_{20}\rho A_{02}-\rho
A_{00}\right) +g_{0}g_{1}^{*}\left( A_{21}\rho A_{02}-\rho A_{01}\right)
\right]  \nonumber \\
&&+F(\omega _{10})N\left[ |g_{1}|^{2}\left( A_{21}\rho A_{12}-\rho
A_{11}\right) +g_{0}^{*}g_{1}\left( A_{20}\rho A_{12}-\rho A_{10}\right)
\right]  \nonumber \\
&&\left. +h.c.\right\},  \label{master1}
\end{eqnarray}
where $F(\pm \omega _{10})=\left[ \kappa +i(\delta \pm \omega
_{10}/2)\right] ^{-1}$.

Obviously, the equation (\ref{master1}) describes the cavity-induced atomic
decay into the cavity mode. The real part of $F(\pm \omega _{10})|g_{i}|^{2}$
represents the cavity-induced decay rate of the atomic excited level $
|2\rangle $ to the ground level $|i\rangle \,$, $(i=0,\,1)$, while the
imaginary part is associated with the frequency shift of the atomic level
resulting from the interaction with the thermal field in the detuned cavity.
The other terms, $F(\pm \omega _{10})g_{i}g_{j}^{*},\,(i\neq j)$, however,
represent the cavity-induced correlated transitions of the atom, i.e., as
the atom emits a photon from the excited level $|2\rangle $ to one of the
ground sublevels, say $|0\rangle $ for example, it drives an absorption of
the same photon on a different transition, $|1\rangle \rightarrow |2\rangle $
, and vice versa, which give rise to the effect of quantum interference.

The effect of quantum interference is very sensitive to the orientations of
the atomic dipoles and the polarization of the cavity mode. For instance, if
the cavity-field polarization is not pre-selected, as in free space, one
must replace $g_{i}g_{j}^{*}$ by the sum over the two possible polarization
directions, giving $\Sigma _{\lambda }g_{i}g_{j}^{*}\propto {\bf d}
_{i2}\cdot {\bf d}_{j2}^{*}$ \cite{agarwal}. Therefore, only non-orthogonal
dipole transitions lead to nonzero contributions, and the maximal
interference effect occurs with the two dipoles parallel. As pointed out in
Refs. \cite{harris,cardimona,agarwal,berman} however, it is questionable
whether there is a isolated atomic system with parallel dipoles. Otherwise,
if the polarization of the cavity mode is fixed, say ${\bf e}_{\lambda }=
{\bf e}_{x}$, the polarization direction along the $x$-quantization axis,
then $g_{i}g_{j}^{*}\propto \left( {\bf d}_{i2}\right) _{x}\left( {\bf d}
_{j2}^{*}\right) _{x}$, which is nonvanishing, regardless of the orientation
of the atomic dipole matrix elements. Actually, by selecting the cavity
polarization, we can in some cases even engineer a system with two parallel
or anti-parallel dipole moments. For example, for an atom with a $
|j,\,m=0\rangle \leftrightarrow |j-1,\,m=\pm 1\rangle $ transition, if we
pre-selected the cavity polarization to the $x$-quantization axis, we will
achieve a scheme with two parallel dipole moments, whereas if the cavity
polarization is pre-selected to the $y$-quantization axis, we will have a
system with two anti-parallel dipole moments.

It is apparent that if $\kappa \gg \delta ,\,\omega _{10}$, the frequency
shifts are negligibly small. Moreover, if we define the cavity-induced decay
rates of the excited level to the ground sublevels as $\gamma _{0}=\kappa
|g_{0}|^{2}/[\kappa ^{2}+(\delta -\omega _{10})^{2}]\simeq
|g_{0}|^{2}/\kappa $ and $\gamma _{1}=\kappa |g_{1}|^{2}/[\kappa
^{2}+(\delta +\omega _{10})^{2}]\simeq |g_{1}|^{2}/\kappa $, the master
equation (\ref{master1}) then reduces to the approximate form

\begin{eqnarray}
\dot{\rho} &\simeq &-i\left[ H_{A},\,\rho \right]   \nonumber \\
&&+\gamma _{0}(N+1)(2A_{02}\rho A_{20}-A_{22}\rho -\rho A_{22})+\gamma
_{0}N(2A_{20}\rho A_{02}-A_{00}\rho -\rho A_{00})  \nonumber \\
&&+\gamma _{1}(N+1)(2A_{12}\rho A_{21}-A_{22}\rho -\rho A_{22})+\gamma
_{1}N(2A_{21}\rho A_{12}-A_{11}\rho -\rho A_{11})  \nonumber \\
&&+2\sqrt{\gamma _{0}\gamma _{1}}(N+1)A_{12}\rho A_{20}+\sqrt{\gamma
_{0}\gamma _{1}}N(2A_{21}\rho A_{02}-A_{01}\rho -\rho A_{01})  \nonumber \\
&&+2\sqrt{\gamma _{0}\gamma _{1}}(N+1)A_{02}\rho A_{21}+\sqrt{\gamma
_{0}\gamma _{1}}N(2A_{20}\rho A_{12}-A_{10}\rho -\rho A_{10}).  \label{maxs}
\end{eqnarray}
This equation is same as that of a $\Lambda $-type three-level atom with two
parallel transition matrix elements in free space \cite{jan}. In other
words, the maximal effect of quantum interference in a $\Lambda $-type atom
can be achieved in a cavity with a pre-selected polarization. Furthermore,
transforming eq. (\ref{maxs}) into the basis: $\left\{ |2\rangle
,\;|S\rangle =\left( \sqrt{\gamma _{0}}|0\rangle +\sqrt{\gamma _{1}}
|1\rangle \right) /\sqrt{\gamma _{0}+\gamma _{1}},\;|A\rangle =\left( \sqrt{
\gamma _{0}}|1\rangle -\sqrt{\gamma _{1}}|0\rangle \right) /\sqrt{\gamma
_{0}+\gamma _{1}}\right\} $, shows that the cavity mode only couples to the
states $|S\rangle $ and $|2\rangle $ with a cavity-induced decay rate of $
\left( \gamma _{0}+\gamma _{1}\right) $, and the asymmetric state $|A\rangle
$ is decoupled from the excited state $|2\rangle $. Interestingly, in the
case of degenerate ground states ($\omega _{10}=0$), the steady-state
solution is highly dependent upon initial conditions of the atom. For
example, if the atom is initially in the asymmetric state $|A\rangle $, it
will stay in the state forever,{\it \ i.e.}, $|A\rangle $ is a complete
trapped state, whereas the steady-state populations are respectively, $\rho
_{22}=N/(2N+1)$, $\rho _{SS}=(N+1)/(2N+1)$ and $\rho _{AA}=0$, if the atom
is initially in either the symmetric state $|S\rangle $ or the excited state
$|2\rangle $. Otherwise, for the atom initially in one of the ground
doublet, $\rho _{22}=N/(4N+2)$, $\rho _{SS}=(N+1)/(4N+2)$ and $\rho _{AA}=1/2
$, where an half population is trapped in the state $|A\rangle $. It is
evident that the existence of the population trapped state and the
dependence of the steady-state population on the initial atomic states
originate from the cavity induced quantum interference..

Our numerical calculations show no trapped state at all in the nondegenerate
case ($\omega _{10}\neq 0$). Nevertheless, the cavity-induced quantum
interference between the two transition paths, $|0\rangle \leftrightarrow
|2\rangle $ and $|1\rangle \leftrightarrow |2\rangle $ gives rise to the
steady-state population inversions and coherence, as shown in Fig. 1, where $
\omega _{10}=2\kappa =200$, $N=20$ and $g_{0}=g_{1}=10$ are taken.$\ $ The
steady-state populations and coherence are highly dependent on the cavity
frequency. The coherence is symmetric with the cavity detuning and reaches
the maximum value at $\delta =0$, while the population differences are
asymmetric. Furthermore, the population inversions may be achieved for
certain cavity frequency. For example, if the cavity frequency is tuned to $
-139.2<\delta <82.3$, the population is inverted between the excited level $
|2\rangle $ and the ground sublevel $|0\rangle $, (i.e., $\rho _{22}>\rho
_{00}$), whereas $\rho _{22}>\rho _{11}$ in the region of $-82.3<\delta
<139.2$. It is clear that $\rho _{22}>\rho _{11}>\rho _{00}$ is achieved in
the region of $-139.2<\delta <0$. The steady-state population inversions and
nonzero coherence manifests the cavity-induced quantum interference \cite
{absen}.

Now we investigate the effects of quantum interference on the Autler-Townes
spectrum $A(\omega )$, by illuminating a weak, frequency-tunable probe field
on such a system. One may predict that, in the absence of the cavity-induced
interference (i.e., no cross transition, associated with $
g_{i}g_{j}^{*}$, is taken into account ),{\it \ }two transition paths, $
|0\rangle \leftrightarrow |2\rangle $ and $|1\rangle \leftrightarrow
|2\rangle $, are {\it independent}, which respectively lead to the higher-
and lower-frequency sidebands of the absorption doublet with respective
linewidths $\gamma _{0}\left( 2N+1\right) +\gamma _{1}\left( N+1\right) $
and $\gamma _{0}\left( N+1\right) +\gamma _{1}\left( 2N+1\right) $. Whereas,
the spectral features may be dramatically modified in the presence of the
cavity-induced interference. Here we only concentrate on the case $\omega
_{10}\sim 2\kappa \gg \gamma _{0},\gamma _{1},N$, so that the doublet is
well resolved. See for example, in Fig. 2 where $\omega _{10}=2\kappa =200$,
$N=20$, $g_{0}=g_{1}=10$ and different cavity detunings are taken, in which
the solid (dashed) lines represent the spectrum in the presence (absence) of
the cavity induced interference. It is clearly shown that, when the cavity
is resonant with the average frequency of the atomic transitions, $\delta =0$
, the interference widens and strengthens the absorption doublet, which is
symmetric, (Fig. 2(a)). Otherwise, it is asymmetric. Rather surprisingly,
probe gain may occur at either the lower- or the higher-frequency sideband,
e.g., the probe field is amplified at the lower-frequency sideband for $
\delta =50$ and $100$, while at the other sideband for $\delta =200$, see in
Figs. 2(b)-2(d) for instance. When the cavity detuning is much larger than
the ground sublevel splitting and the cavity linewidth, $\delta \gg \omega
_{10},2\kappa $, the effect of the cavity induced interference is negligible
small so that the absorption spectrum is virtually same as that without
interference (we show no figure here).

It is well known that the probe absorption of multi-level atoms is
attributed to population difference between two dipole transition levels and
coherence between two dipole forbidden levels, and either the inverted
populations or the coherence can lead to probe gain. As demonstrated in Fig.
1, the population between the two transition levels $|2\rangle $ and $
|1\rangle $ is inverted in the region of $-82.3<\delta <139.2$. Therefore,
the gain at the lower-frequency sideband stems from the cavity-induced
steady-state population inversion between $|2\rangle $ and $|1\rangle $ for $
\delta =50$ and $100$, whereas the cavity-induced coherence between the two
dipole-forbidden excited sublevels $|0\rangle $ and $|1\rangle $ must be the
origin of the gain at the higher-frequency one in the case $\delta =200$.

In summary, we have shown that maximal quantum interference can be achieved
in a $\Lambda $-type atom coupled to a single-mode, frequency-tunable cavity
field at finite temperature, with a pre-selected polarization in the bad
cavity limit. The cavity-induced interference may give rise to the
population trapping and inversions, and the probe gain at either sideband of
the Autler-Townes doublet, depending upon the cavity resonant frequency, the
ground level splitting and the mean number of thermal photons. The gain
occurring at different sidebands has the various origin: in the case of $
\delta >0$, the higher-frequency gain is due to the nonzero coherence, while
the lower-frequency one is attributed to the population inversion. As shown
in Refs: \cite{EIT,nar,scully} that an apply laser coupling to multilevel
atoms may result in the steady-state coherence and population inversions. We
here present an another scheme whereby they can be generated by the
cavity-induced interference.

We should emphasize that there are no special restrictions on the atomic
dipole moments in our system, as long as the polarization of the cavity
field is pre-selected, and that the effects of the cavity-induced
interference occur over ranges of the parameters, and are profound when the
ground level splitting is the same order of the cavity linewidth and the
mean number of thermal photons $N\gg 1$, which may make its experimental
observation feasible.

\acknowledgments

I greatly acknowledge conversations with Z. Ficek and S. Swain.

\begin{figure}[tbp]
\caption{The steady-state population differences and coherence vs the cavity
detuning, for $g_{0}=g_{1}=10,\,\kappa=100,\,\omega_{10} =200$ and $N=20$.
The solid, dashed and dot-dashed lines respectively represent $(\rho_{22}
-\rho_{00})$, $(\rho_{22} - \rho_{11})$ and Re($\rho_{01}$).}
\label{fig1}
\end{figure}

\begin{figure}[tbp]
\caption{Absorption spectrum $A(\omega)$ vs the scaled frequency $
\omega=(\omega_p -\omega_0)$, where $\omega_p$ is the frequency of the probe
field, for $g_{0}=g_{1}=10$, $\kappa =100,\,\omega _{10}=200,\,N=20$, and $
\delta =0,\,50,\,100,\,200$ in (a)--(d), respectively. The solid curves
represent the spectrum in the presence of the cavity-induced interference,
whilst the dashed curves are the spectrum in the absence of the
interference. }
\label{fig2}
\end{figure}


\begin{references}
\bibitem{absen}  Noting that in the sdeady state, $\rho _{00}=\rho
_{11}=(N+1)/(3N+2)$, $\rho _{22}=N/(3N+2)$, and $\rho _{01}=0$ in the
absence of the interference.
\end{references}
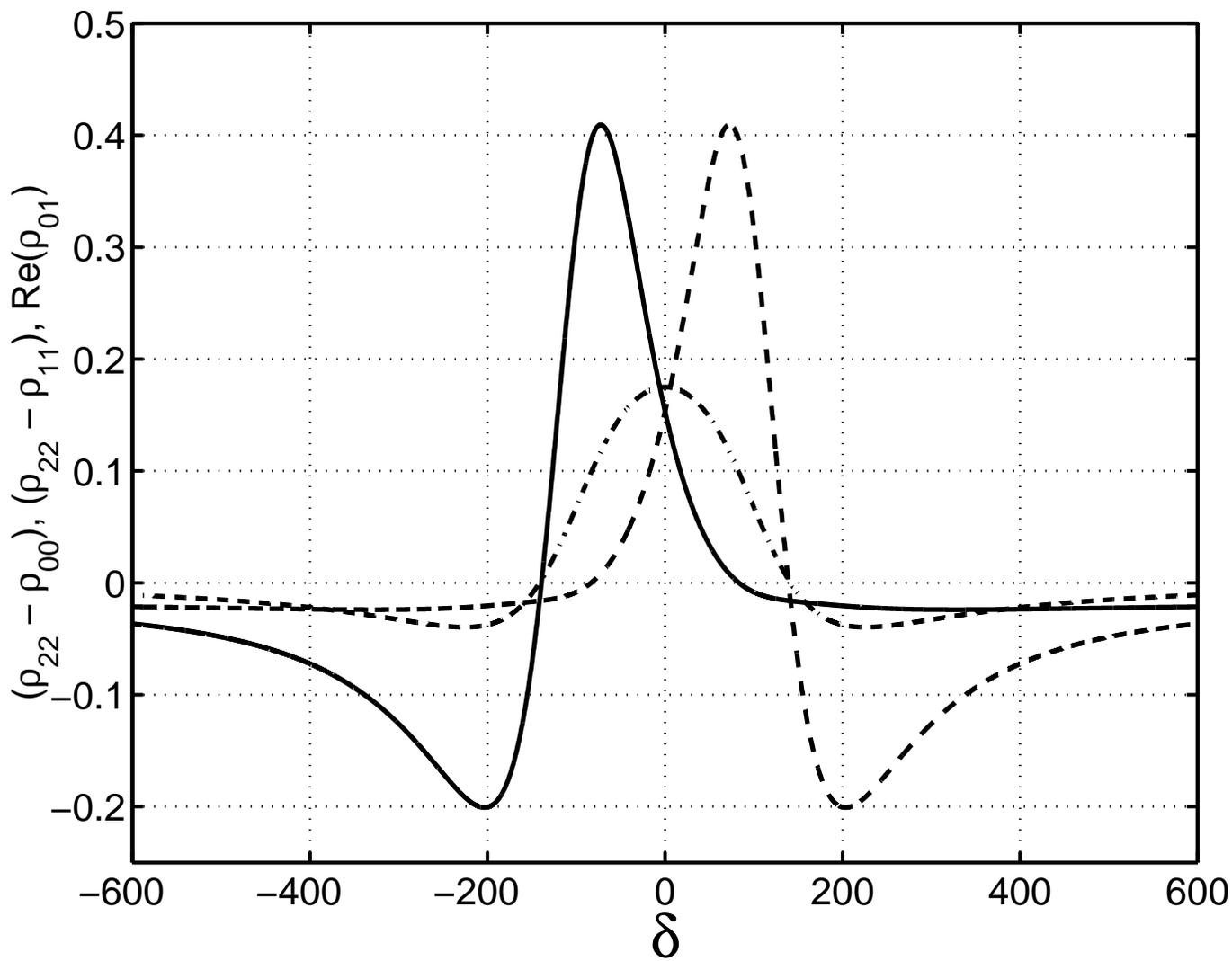
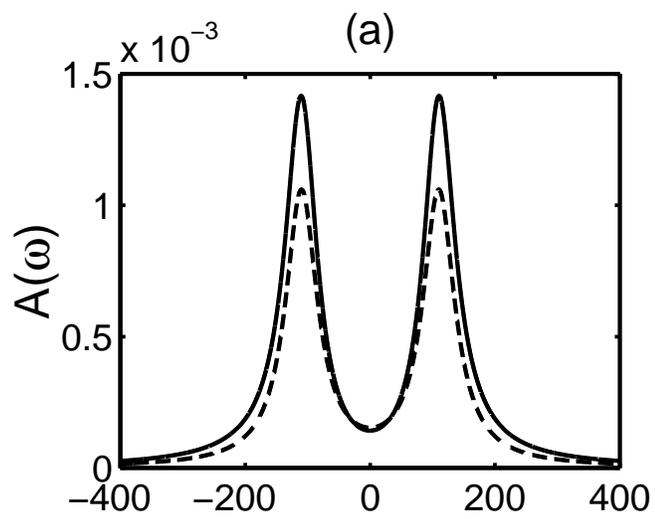
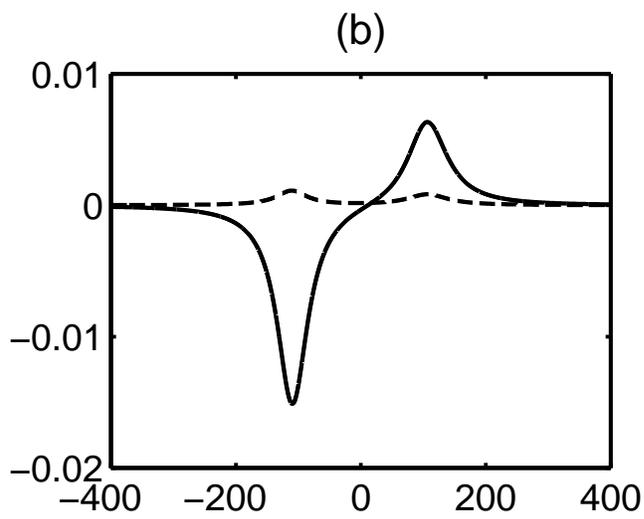
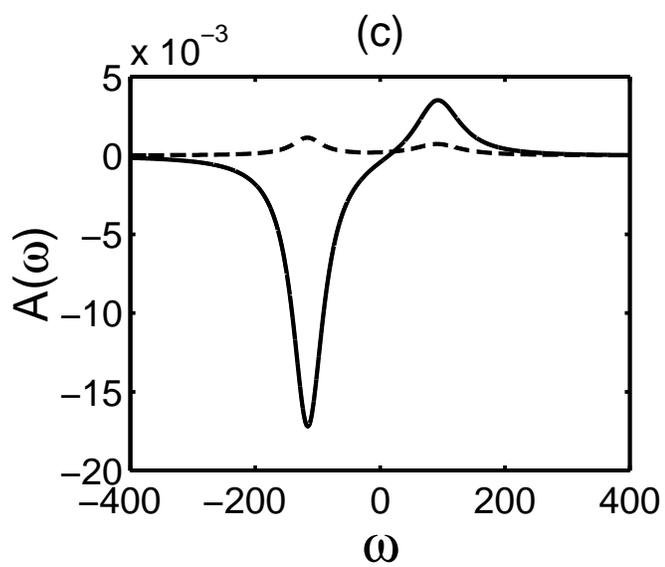
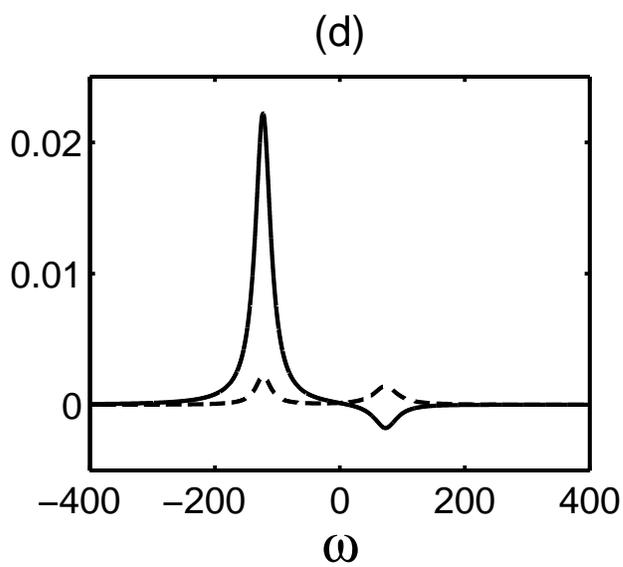
\end{document}